# Pirus : A Web Based File Hosting Service with Object Oriented Logic in Cloud Computing

Dimitrios Kallergis[1] , Konstantinos Chimos[2], Vizikidis Stefanos[3], Theodoros Karvonunidis[4], Christos Douligeris[5]

Department of Informatics, University of Piraeus

Piraeus, Greece

{D.Kallergis, himosk, cdoulig}@unipi.gr

svizkidis@gmail.com, tkarv@otenet.gr

**Abstract**:

In this paper a new Web-based File Hosting Service with Object Oriented Logic in Cloud Computing called Pirus was developed. The service will be used by the academic community of the University of Piraeus giving users the ability to remotely store and access their personal files with no security compromises. It also offers the administrators the ability to manage users and roles. The objective was to deliver a fully operational service, using state-of-the-art programming techniques to enable scalability and future development of the existing functionality. The use of technologies such as .NET Framework, C# programming language, CSS and jQuery, MSSQL for database hosting and the support of Virtualization and Cloud Computing will contribute significantly in compatibility, code reuse, reliability and reduce of maintenance costs and resources .

The service was installed and tested in a controlled environment to ascertain the required functionality and the offered reliability and safety with complete success.

The technologies used and supported, allow future work in upgrading and extending the service. Changes and improvements, in hardware and software, in order to convert the service to a SaaS (Software as a Service) Cloud application is a logical step in order to efficiently offer the service to a wider community. Improved and added functionality offered by further development will leverage the user experience.

I. Introduction

New trends in academia and industry operations demand the capability of sharing-from-anywhere digital material, as well as retrieving multimedia and documents from multiple repositories over the Internet. In this paper, we introduce the implementation of a file hosting platform using object oriented logic in a cloud computing environment. This platform may be prospectively used in collaborative processes taking place in multiple working environments.

Our goal is to implement a fully functional file hosting service over the Net using innovative programming and architectural techniques. This is the first step of improving existing tools currently that may be in use in multiple working environments; such as academic institutes.

There have been used object oriented techniques during the design and implementation phase of this platform; such as .NET Framework ecosystem that facilitates source code reuse in order to provide rapid and safe software application development. Moreover, it supports different production environments such as the Internet and Organisations' Intranets providing functionality and customisation of the service. It also adopts and supports methods and techniques from the Cloud ecosystem [1], [2], providing extensibility and elasticity; our aim is to aggregate multiple users from multiple environments or organisations.

The platform was installed and tested in a controlled environment to ascertain the required functionality and the offered reliability and safety with complete success. This platform's aim is to be the corner stone of a collaboration environment in the industry or in the academic learning and research processes.

This paper presents the offered functionality, the programming and architectural tools and techniques that were used and are provided by the service, as well as future implementation and add-on work for it.

## II. The Web-Based File Hosting Platform

### A) Functionality

The implemented file hosting platform is the main part of a web-based service for collaborative purposes. Whoever uses this service is able to save digital material, to create files and folders within the file system, as well as to share these data using desktop or mobile computing machinery.

The platform also provides the creation of multiple users or multiple groups of users that will share their data. The users or their groups may have different rights for accessing these data; this approach gives us the ability of offering the service into human environments with multiple accessing digital rights (e.g. supervisor/employee, teacher/student, *etc.*).

The administration of the platform is also a vital issue. This vital part is implemented through a special gateway as the provided tools allow in real time the creation/update/delete of users, role assignment to them, as well as their attributes' manipulation; such as user quota.

The file hosting platform was implemented as a part of an interworking service over the Cloud using functionalities from the *.NET 4 Framework*, and fully providing compatibility when accessing from a desktop computer or a mobile smart device. The access to the service is agnostic to the web page browser technology and operating system as it does not require a special software agent or application for this purpose.

### B. Architectural and development issues

#### 1) Architectural issues

Object oriented logic was used to design this project in order to minimize the software development time, to have a homogenous total of source code that can be easily reused in the future and to ensure the safe execution of code itself. During the design process, the members of our team had the opportunity to consider different use cases, as well as to conduct sequence diagrams for the operation of the platform and the service respectively.

Additionally, a relational database system was used with cloud-ready functionality. This system provides scalability and elasticity according to the service's demand and can be easily deployed in the Cloud.

We also used *~okeanos* [3], which is an infrastructure as a service (IaaS) project provided by the Greek Research and Technology Network (GRNET S.A.). Okeanos was used in order to facilitate the server side of the file hosting platform within the Cloud ecosystem. Under this, we managed to provide our software as a service (SaaS) solution with elasticity and scalability features through *{a}* the immediate re-specifying of the technical resources, *{b}* the improvement of the reliability in a case of physical disaster, *{c}* the ease in software maintenance because the platform operates in distributed mode and *{d}* the cost minimization in terms of hardware's maintenance, operational effort and software development time.

## 2. Development issues

We implemented our source code with *C# programming language* within the *.NET Framework* [4]. We also used *MS Intermediate Language* for rapid software development. Code reuse was feasible through *{a} Master Pages, {b} User Controls* and *{c} Web Forms* from ASP.NET essentials. *Cascading Style Sheets* [5] were used for designing web pages with dynamic format and view, as well as *jQuery* [6] for ultimate compatibility for every web browser. Finally, *MSSQL* database system was almost binding for us because of its functionality and interaction with the whole .NET Framework.

3. **Implementation**

The basic element of our file hosting service is the view and interaction of digital material from an individual. The user should be able to see the file and folder structure that has created and through this to be able to update its own digital content.

Our system has taken into consideration security issues for the service connection, as well as its personalized operation. The .NET Framework offers several ways to implement this in a software level; we used *{a} Membership*, to create and handle users, *{b} Roles*, to administrate different roles within the working environment (*i.e.* supervisor/employee, teacher/student), *{c} Profile,* to save user data in the database system.

1. Use Cases

We provide several UML diagrams examining different operations of our system.

*a)* Login subsystem

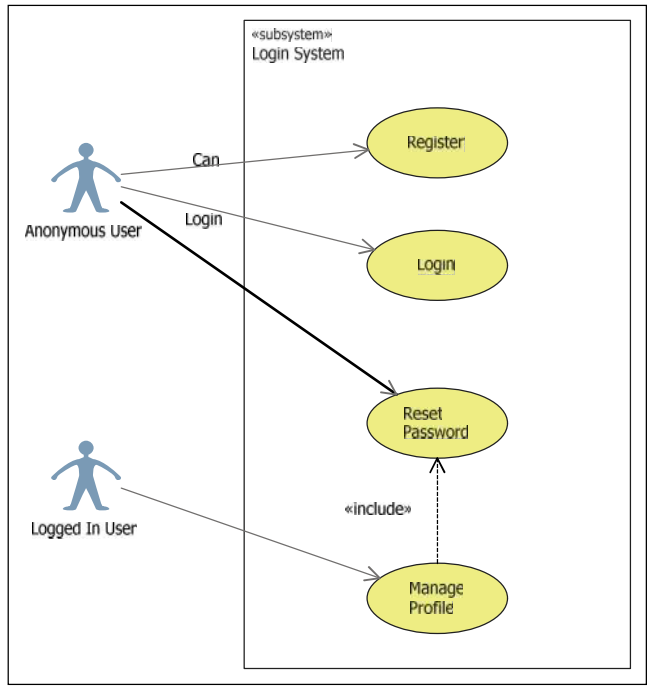

Figure 1. Login Subsystem

2. Service central functionality subsystem

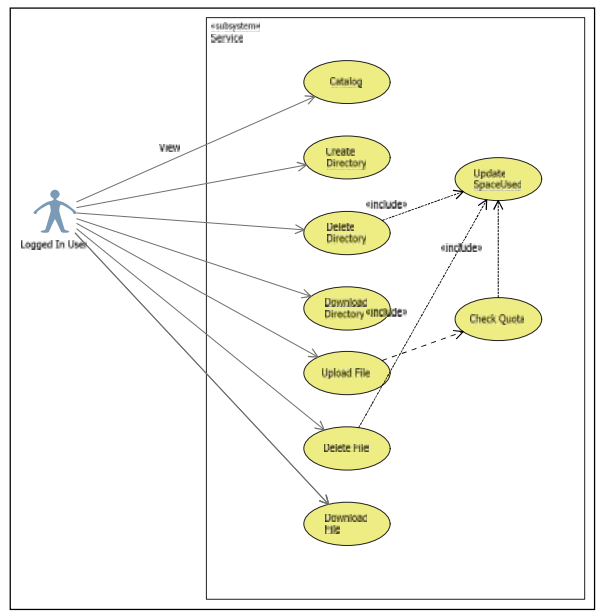

Figure 2. Service central functionality subsystem

3. *Administration subsystem*

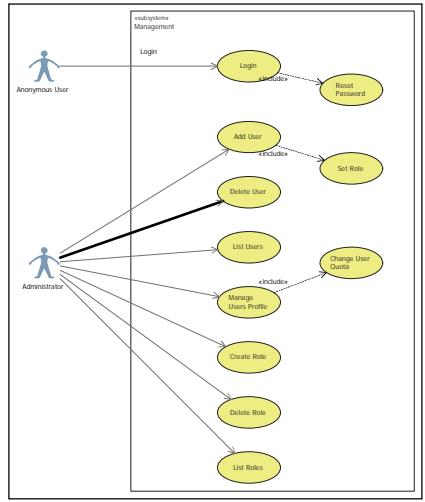

Figure 3. Administration subsystem

Conclusions and Future Work

A Web-based File Hosting Service was presented for the purpose of this paper. Our aim was to design and develop a fully functional service which will provide the opportunity for extending its capabilities, using tools that are being involved in multiple working environment processes.

We tested this platform along with the *~Okeanos*; the IaaS project provided by GRNET. Most of the users that took part in the system's testing period were students. Their behavioral feedback gave us the motive to think about going our project further.

Our service may be the basis of future implementation for internetworking applications with more specific use; e.g. an E-learning system. We present some ideas that may help for this purpose.

a) File versioning

There is a deficit in known file systems; that is the absence of keeping versions of any file change. We propose that this feature should be a part of a future implementation; especially if this system will be used for educational purposes. The user (e.g. teacher/student) should be able to add some comments to any file version and these data will be saved as metadata information.

b) Sharing digital content with other users or groups

Users or group of users should be able to share their digital content. This feature will be implemented taking for granted files/folders access rights in a file system.

**File preview**

An individual should be able to preview documents, spreadsheets or files in portable format (.pdf) without having the obligation to download it to its physical storage medium.

c)  Files and Folders Logical Links

Assuming that a future implementation will be a part of an E-learning system [7], we must point that the logical wrap-up of files within folders will be a hard problem; because of the total amount of data being stored and transferred. Another solution should be proposed; let us say that files and folders are connected to each other using a root-object. This root-object's leafs are files and folders. In that case the user will have the opportunity to access and retrieve any data that has a logical connection with him; this connection will be made through the root-objects.

d)  Distributed File System

Future implementations may offer the ability of using distributed file systems; such as CIFS, NFS, AFS. Further examination will also be made upon this issue regarding handling data separately; physical data and metadata into different nodes [8]. Other experimental implementations should also be considered [9].